%% file: PTL2019_arxiv_V2.tex
\documentclass[journal]{IEEEtran}
\usepackage{amsmath,amssymb,amscd,latexsym,dsfont}
\usepackage{comment}
\usepackage{color}
\usepackage{subfigure}
\usepackage{enumerate}
\usepackage{enumitem}
\usepackage{stfloats}
\usepackage{psfrag}
\usepackage{setspace}
\usepackage{cite}
\usepackage{balance}
\usepackage{enumitem}
\usepackage{tikz}
\usepackage{pgfplots,amsfonts}
\usepackage{pgfplotstable}
\usetikzlibrary{shapes}
\usetikzlibrary{spy}
\usetikzlibrary{fit}
\usetikzlibrary{shapes.multipart}
\usetikzlibrary{positioning}
\input{myFigures.tex}

\input{myStyles.tex}

\input{myMacros.tex}

\input{myPackages.tex}
\pgfplotsset{compat=1.14}

\newcommand{\TypeOne}{8D-2048PRS-T1 }
\newcommand{\TypeTwo}{8D-2048PRS-T2 }

\IEEEoverridecommandlockouts

\title{ {Eight-dimensional  Polarization-ring-switching Modulation Formats}}

\author{Bin Chen,~\IEEEmembership{Member,~IEEE}, Chigo Okonkwo,~\IEEEmembership{Senior Member,~IEEE}, Hartmut Hafermann \IEEEmembership{Senior Member,~IEEE}, \\ and Alex Alvarado,~\IEEEmembership{Senior Member,~IEEE}.
\thanks{B. Chen is with  the School of Computer and Information Engineering, Hefei University of Technology, Hefei, China.
B. Chen is also with Eindhoven University of Technology, 5600 MB, Eindhoven, The Netherlands  (e-mail:~chen.bin.conan@gmail.com).}
\thanks{C. Okonkwo  is with the Institute for Photonic Integration, Eindhoven University of Technology, Eindhoven 5600 MB, The Netherlands (e-mail:~c.m.okonkwo@tue.nl).}
\thanks{A. Alvarado is with the Information and Communication Theory Lab, Signal Processing Systems Group, Department of Electrical Engineering, Eindhoven University of Technology, 5600 MB, Eindhoven, The Netherlands (e-mail:~a.alvarado@tue.nl).}
\thanks{Hartmut Hafermann is with the Optical Communication Technology Lab, Paris Research Center, Huawei Technologies France SASU, 92100 Boulogne-Billancourt, France (e-mail:~hartmut.hafermann@huawei.com).}
\thanks{Research supported by Huawei France through the NLCAP project. The work of B. Chen is supported  by the National Natural Science Foundation
of China (NSFC) under Grant 61701155 and Fundamental Research Funds for the Central Universities under Grant JZ2019HGBZ0130. The work of A. Alvarado is supported by the Netherlands Organisation for Scientific Research (NWO) via the VIDI Grant ICONIC (project number 15685).}
}

\begin{document}
\maketitle

\begin{abstract}
We propose two 8-dimensional (8D)  modulation formats (8D-2048PRS-T1  and 8D-2048PRS-T2) with a spectral efficiency   of 5.5 bit/4D-sym, where the 8 dimensions are obtained from two time slots and two polarizations. Both formats provide a higher tolerance to nonlinearity by selecting  symbols with nonidentical states of polarization (SOPs) in two time slots.
The performance of these novel 8D modulation formats is assessed in terms of the effective signal-to-noise ratio (SNR)  and normalized generalized mutual information.
\TypeOne is more suitable for high SNRs, while \TypeTwo is shown to be more tolerant to nonlinearities.
 {A sensitivity improvement of at least 0.25~dB is demonstrated
by maximizing normalized generalized mutual information
(NGMI). For a long-haul nonlinear optical
fiber transmission system, the benefit of mitigating the nonlinearity is demonstrated and  a reach
increase of $6.7\%$ (560~km) over time-domain hybrid four-dimensional two-amplitude eight-phase
shift keying (TDH-4D-2A8PSK) is observed.}
\end{abstract}
\begin{IEEEkeywords}
Achievable information rates, fiber nonlinearity, generalized mutual information, multidimensional modulation.
\end{IEEEkeywords}

\vspace{-1em}
\section{Introduction}\label{Sec:Introduction}
\IEEEPARstart{F}{iber} nonlinearities are considered to be one of the limiting factors for achieving higher information rates in coherent optical transmission systems \cite{EssiambreJLT2010}. 
Advanced modulation formats with geometric  and probabilistic shaping have been extensively explored with the aim of increasing achievable information rates (AIRs) \cite{Karlsson:09,AgrellJLT2009, TobiasJLT16}. Meanwhile,   signal shaping has also been  considered to mitigate the effects of fiber nonlinearities \cite{Shiner:14,Kojima2017JLT,El-RahmanJLT2018,BendimeradECOC2018,BinChenJLT2019}.

Polarization multiplexing (PM) naturally allows modulation on a four-dimensional (4D) space, which has the potential to increase achievable information rates when the modulation is truly designed in 4D. Conventional PM-formats such as PM-$M$QAM, however, are only optimized per two dimensions independently, and thus, do not exploit all the available degrees of freedom. Several power-efficient modulation formats have been proposed using sphere-packing and lattice constructions in 4D and 8D space \cite{AgrellJLT2009,Karlsson:09,KoikeAkinoECOC2013,Millar:14}. These designs, however, aim at optimizing the minimum Euclidean distance (ED) of the constellation, and thus, they are optimum only for asymptotically high SNR, in the linear additive white Gaussian noise (AWGN) channel, and for uncoded metrics such as symbol- and bit-error probability only \cite[Sec.~IV-A]{AlexTIT2018}.
Some of these multidimensional (MD) modulation formats were also shown to give high mutual information (MI), but are not well-suited for coded systems based on a bit-wise decoder such as bit-interleaved coded modulation (BICM), i.e., their generalized mutual information (GMI) is quite low \cite{Alvarado2015_JLT}.

MD constant-modulus modulation formats have been proposed \cite{Chagnon:13,ReimerOFC2016,Kojima2017JLT} to mitigate the nonlinear interference. One example of this is the 4D 64-ary polarization-ring-switching (4D-64PRS) format we recently proposed in \cite{BinChenJLT2019}.  4D-64PRS was shown to outperform other modulation formats at SE of 6 bit/4D-sym by jointly optimizing the coordinates and labeling. 8D modulation formats have twice as many degrees of freedom, and thus, can  improve the AIRs and nonliearity tolerance. The 8 dimensions can be obtained by two frequencies \cite{ErikssonECOC2013} or two consecutive time slots \cite{KoikeAkinoECOC2013,Shiner:14}.

 Our work builds upon the polarization-balancing concept, proposed for a spectral efficiency (SE) of 2 bit/4D-sym  in \cite{Shiner:14}. This concept was further investigated in terms of the SE and nonlinearity tolerance trade-off in 
 \cite{El-RahmanJLT2018,Bendimerad:18}. All the previous works using this concept only consider PM-QPSK with added constraints, and thus, only 8D  formats at SE below 4 bit/4D-sym were considered. Generalizing those formats to higher SEs is nontrivial, specially when both the constellation and its binary labeling are taken into account.

 In this paper, we  propose an approach to construct two nonlinearity-tolerant modulation formats with a SE of 5.5 bit/4D-sym. The formats are based on set-partitioning 4D-64PRS in two consecutive time slots. The first format is suitable for a high code rate coded modulation system. The second is well-suited for lower code rates and also exhibits higher nonlinearity tolerance. 
 Numerical simulations demonstrate increased nonlinearity tolerance and transmission reach increase with respect to other modulation formats.

\vspace{-0.5em}
\section{8D Polarization-ring-switching Formats}\label{sec:design} 

\begin{figure}[!tb]
	\centering
	\scalebox{0.9}{
	\includegraphics[scale=1]{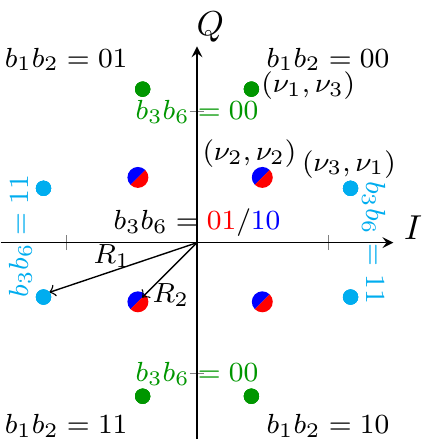}\hspace{0.5em}
	\includegraphics[scale=1]{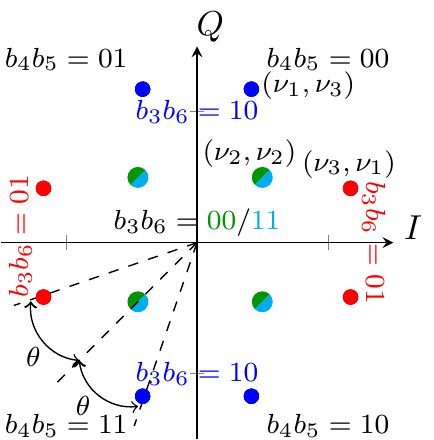}
	}
	\vspace{-0.5em}
	\caption{2D-projections of 4D-64PRS and its binary labeling. The rings are given by $R_1^2={\nu_1^2+\nu_3^2}$ and $R_2^2=2\nu_2^2$.}
	\label{fig:4D_64_modulation_label}
\vspace{-1.5em}
\end{figure}

In optical transmission systems, the performance of a given modulation format is  determined by its tolerance
to both nonlinear interference arising from the Kerr effect, and accumulated amplified spontaneous emission  noise.
Therefore, designing modulation formats which increase the AIRs in the presence of linear and nonlinear impairments is crucial. In \cite{BinChenJLT2019}, we designed the 4D-64PRS format with SE 6~bit/4D-sym, which has a constant modulus and an optimized binary labeling. 4D-64PRS provides excellent linear gain and nonlinear gain with respect to other modulation formats at the same SE. The structure and binary labeling of 4D-64PRS is shown in Fig.~\ref{fig:4D_64_modulation_label}.
The bits $b_1,b_2,b_4,b_5$ determine the two 2D quadrants while $b_3,b_6$ determine the actual transmitted symbol.

\begin{figure}[!tb]
    \centering
    \begin{tabular}{c}
\scalebox{0.83}{    
     \begin{subfigure}[\TypeOne Left: time slot 1. Right: time slot 2.]{
     \includegraphics[width=0.235\textwidth]{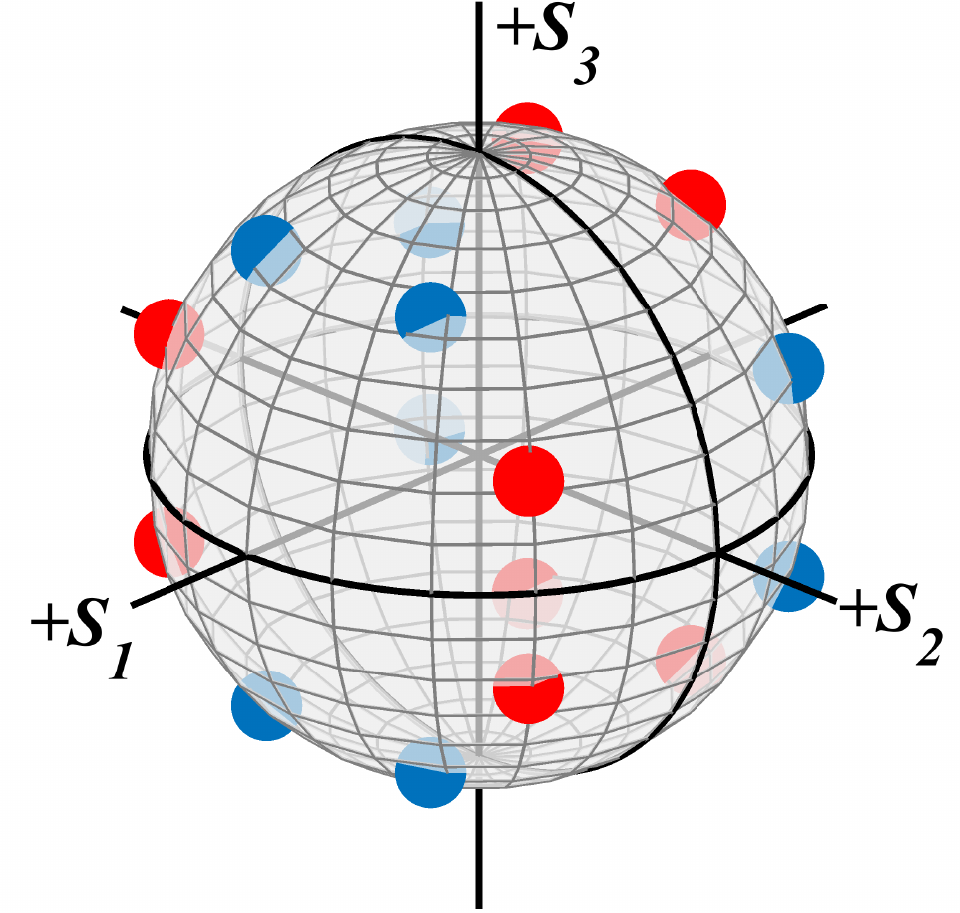}\hspace{1.1em}
     \includegraphics[width=0.235\textwidth]{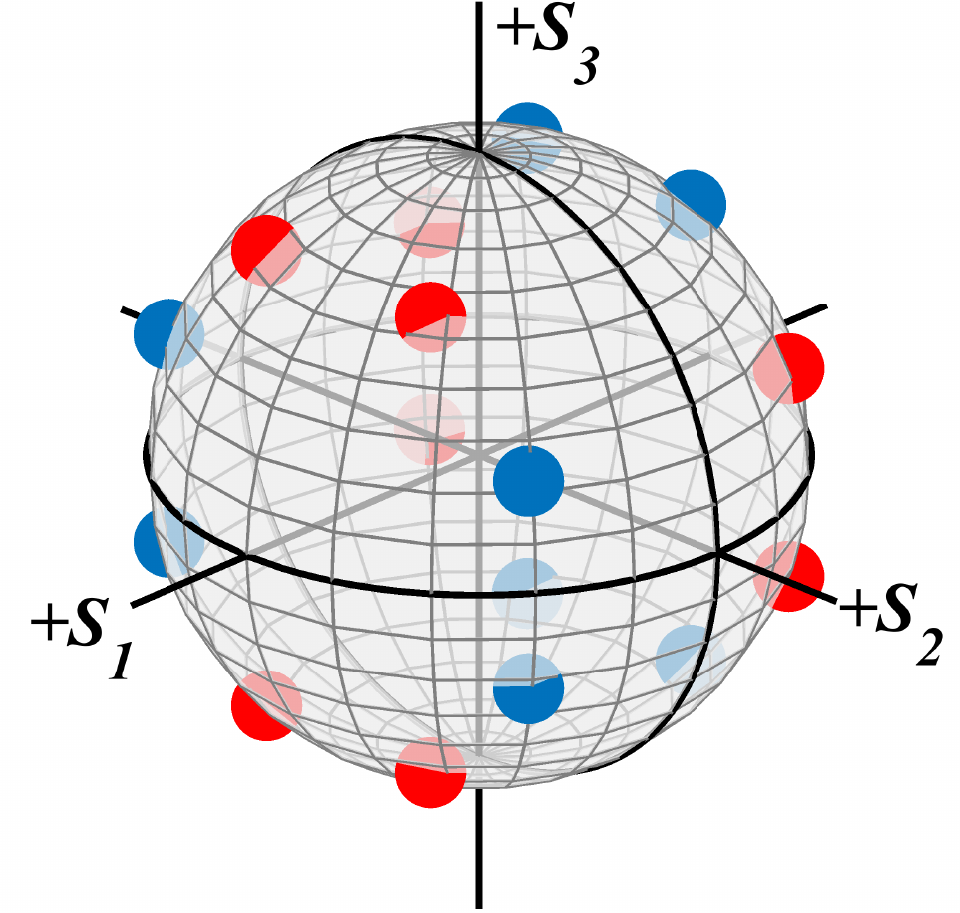}
     }
   \end{subfigure}
   }
   \\
\scalebox{0.83}{ 
    \begin{subfigure}[\TypeTwo Left: time slot 1. Right: time slot 2. ]{
     \includegraphics[width=0.25\textwidth]{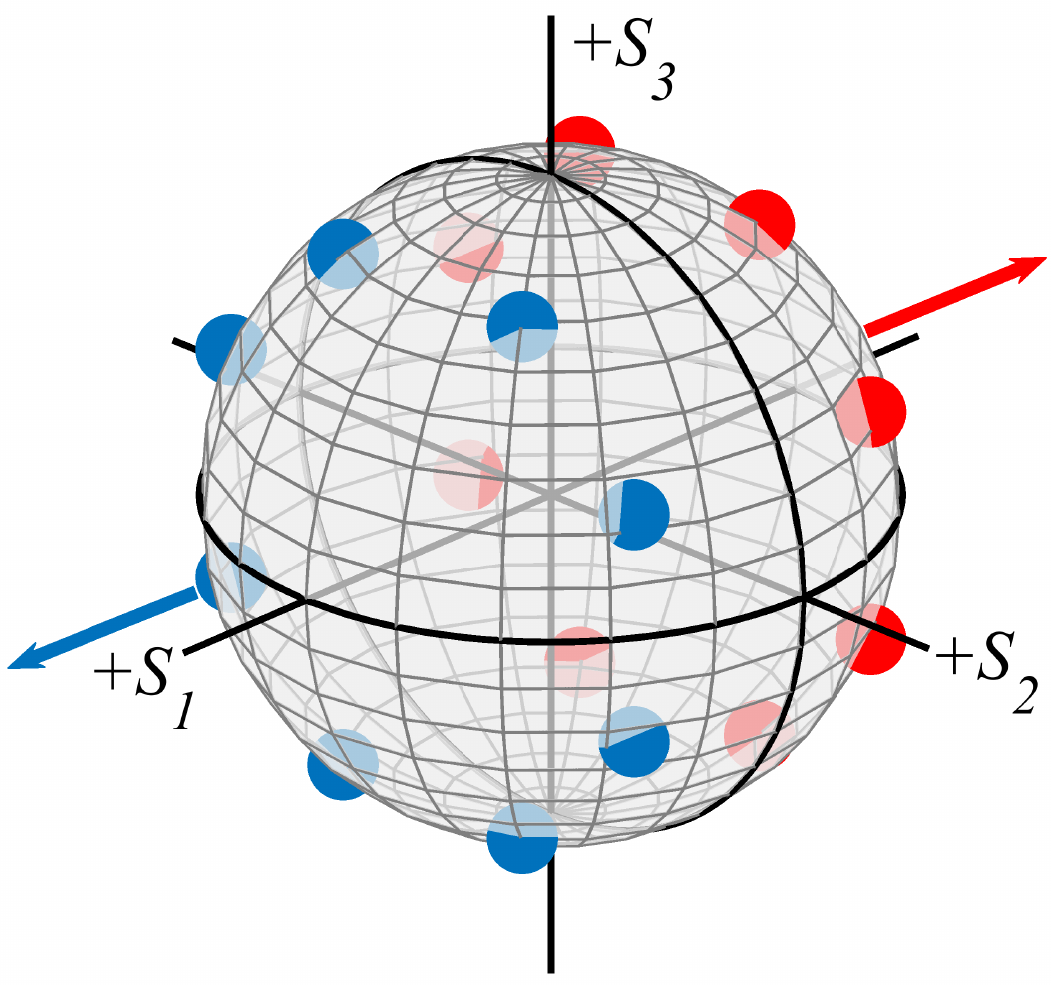}\hspace{0.5em}
     \includegraphics[width=0.25\textwidth]{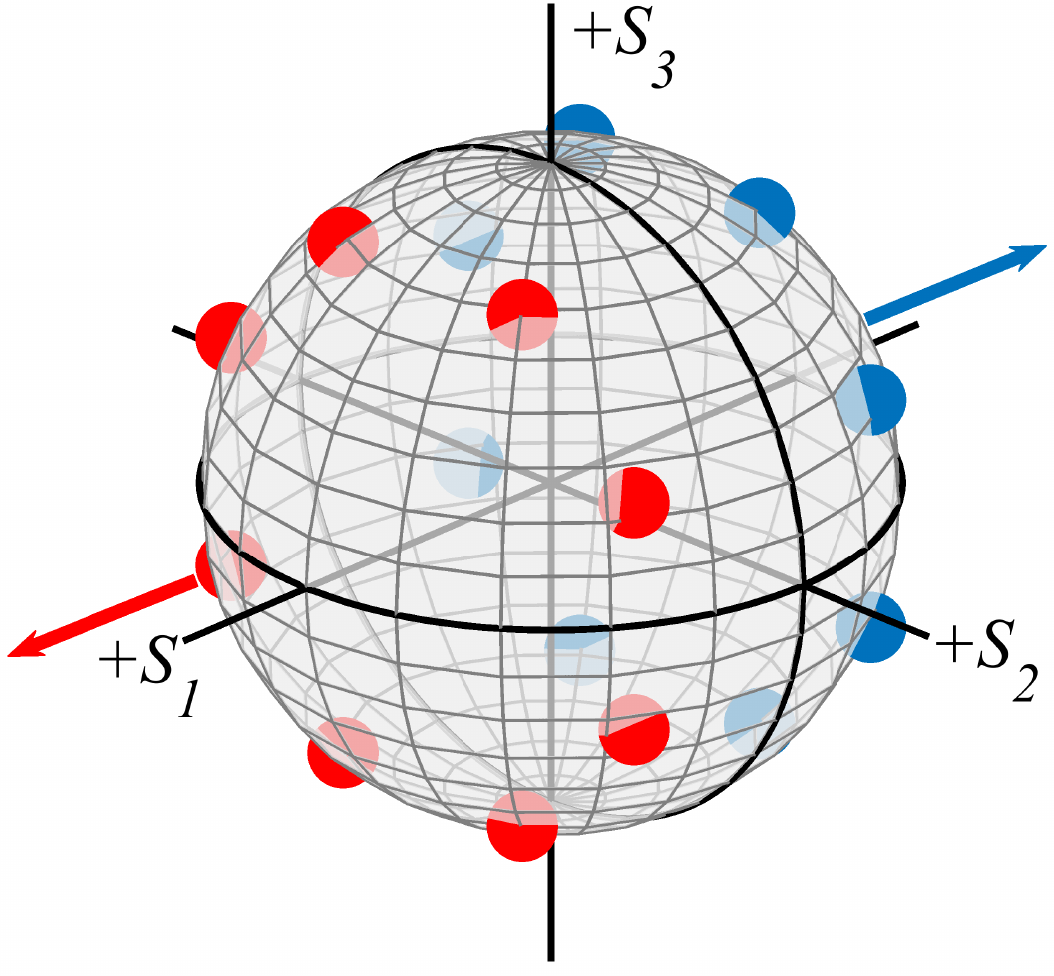}
     }
\end{subfigure}
}
\end{tabular}
\vspace{-0.75em}
\caption{Stokes representation of the designed 8D format for two consecutive time slots. When colors are not considered, all four figures correspond to the Stokes representation of 4D-64PRS.}

    \label{fig:8D-2048PRS}
\vspace{-1em}
\end{figure}

Let $\bS=[S_1,S_2,S_3]$ denote the Stokes vectors
with $S_1= |X|^2-|Y|^2$, $S_2=2\Re\{XY^*\}$, and $S_3=2\Im\{XY^*\}$, and where $X$ and $Y$ are complex numbers representing the constellation symbols in $\text{x}$- and $\text{y}$-polarization, resp. The symbols 4D-64PRS result in 16 distinct states of polarization (SOPs) and    {have} a constant modulus ($\|\bS\|=1$). This is shown in Fig.~\ref{fig:8D-2048PRS} (ignoring the colors).
If the 4D-64PRS format was to be used in two consecutive time slots ($T_1$ and $T_2$), there are $2^{12}=4096$ 8D symbols as a set $\mathcal{X}\in\mathbb{R}^8$, which can be represented by 12 bits $b_1, b_2, \ldots, b_{12}$. In this paper, we are interested in designing formats with a SE of  {5.5~bit/4D-sym (11~bit/8D-sym)}, and thus, we will use $b_{12}$ as parity bit to effectively remove $2048$ out of the $4096$ symbols.

In order to achieve better performance for optical fiber communication system, we  {designed} 8D modulation formats with a better sensitivity and a high nonlinearity tolerance by selecting  symbols  with  larger minimum Euclidean distance and
smaller degree of polarization (DOP) in consecutive time slots.
The DOP for $i$th transmitted 8D symbol is defined as  $p_i=\frac{{||\bS_{t_1}+\bS_{t_2}||}}{|X_{t_1}|^2+|Y_{t_1}|^2+|X_{t_2}|^2+|Y_{t_2}|^2}$, where $0\leq p_i \leq 1$, $t_1$ and $t_2$ indicate time slot 1 and time slot 2. 
It is known that the worst symbols for nonlinearity tolerance are polarization identical (PI)  symbols with zero DOP ($p=0$), which has identical SOPs. 
 {Therefore, we first avoid all the strongest cross polarization
modulation (XPolM)-inducing PI symbols contained in 4D-64PRS and then jointly consider SOP and Euclidean distance to select 2048 polarization nonidentical
symbols ($p<1$) in 4D-64PRS constellation set for two SNR regimes: high-SNR and medium-SNR.}
We obtained two types of 8D modulation formats with  {5.5~bit/4D-sym}. One overhead bit is employed to choose points from the set $\mathcal{X}$ and can be obtained by the following methods:
\begin{itemize}[leftmargin=1ex]
    \item Type 1: 
$b_{12}$ is a parity bit of single-parity-check code to protecting all information bits, which is an exclusive or (XOR) of all the bits $b_1, b_2, \cdots, b_{11}$.  {In this case,  the nearest neighboring  {symbols} are removed to maximize minimum ED, which perform better at higher SNR.} 
The parity bit $b_{12}$ can be obtained as
$ {\overline{b}_{12}= {b_1\oplus b_2\oplus \cdots\oplus b_{10}\oplus b_{11}},}$
where $\oplus$ and $\overline{\cdot}$ denote the modulo-2 addition and negation,
respectively.
\item Type 2: $b_{12}$ is used to protect only the least significant
bits, which are $b_3$, $b_6$ and $b_9$.  {In this case, the modulation will be good for medium SNR. In addition, it has more polarization balanced points in two time slots.}  
The parity bit $b_{12}$ can be obtained as 
$\overline{b}_{12}={b_3\oplus b_6\oplus b_9}.$
\end{itemize}

Fig.~\ref{fig:8D-2048PRS}  shows the relationship of SOPs for transmitted symbols in  two consecutive time slots for two designed 8D modulation formats. 
The  color  coding  scheme  used  {in} Fig.~\ref{fig:8D-2048PRS} shows the SOP constraint we imposed on the formats. When a  blue point   is transmitted  in  the  first time slot, only red points   {are} used in the second time slot. No PI symbols with $p=0$ are allowed in both of these two 8D modulation formats. 

 \begin{table}[!tb]\caption{Comparison of
different modulation formats with SEs  {of} 5.5~and~6.0~[bit/4D-sym].}
\label{tab:property}
\vspace{-0.5em}
\centering
\scalebox{0.7}
{
\begin{tabular}{c|c|c|c|c|c}
\hline
\hline
&SE& $d^2_E$ & $\alpha$ & $\beta$ & Modulus \\
\hline
\hline
PM-8QAM & 6 & 0.84 & 1  &0.70 &Not Constant \\  
\hline
4D-2A8PSK \cite{Kojima2017JLT}& 6 & 0.88 & 1  &0.65 &Constant \\   
\hline
4D-64PRS \cite{BinChenJLT2019}& 6 & 0.66 & 1 &0.65 & Constant\\
\hline 
\TypeOne& 5.5& 1.15  & 0.96  & 0.64 &Constant\\   
\hline
\TypeTwo& 5.5& 0.76 & 0.87  & 0.55 &Constant\\   
\hline
\end{tabular}
} 

\vspace{-1.5em}
\end{table}

To inform our intuition on design features that influence linear and nonlinear performance, we list the properties of five modulation formats in Table \ref{tab:property} for comparison. 
The squared minimum Euclidean distance is  denoted as $d^{2}_E$.  
In addition to constant modulus, we propose two performance metrics for evaluating modulation-dependent nonlinear interference: the maximum DOP and the average DOP, which are calculated for  all the possible $M$ transmitted symbols in two consecutive time  {slots} for a given modulation format.
The maximum DOP is defined as $\alpha=\max_{i\in\{1,2,\ldots,M\}} p_i$ and the average DOP is denoted as $\beta=\frac{1}{M}\sum_{i=1}^{M}p_i$. 
A larger $d^{2}_E$ should  result in better  {linear} sensitivity, while smaller $\alpha$ and $\beta$  should in principle result in higher  nonlinear noise tolerance.
Based on these properties,   {the} two 8D modulation formats should be better than   {the} other   {three} modulation formats for both linear and nonlinear regime, which will be shown in Sec.~\ref{Sec:Simulation}.

\vspace{-0.5em}
\section{Performance Evaluation}\label{Sec:Simulation}
Here we compare the performance of  {four} different modulation formats: PM-8QAM, 5.5b4D-2A8PSK\footnote{{The constellation 5.5b4D-2A8PSK is generated by using 5b4D-2A8PSK and 6b4D-2A8PSK from \cite{KojimaOFC2017} with optimized ring ratio in a time-domain hybrid way with a 1:1 ratio. }}, and the two proposed 8D-2048PRS formats.  {We use the PM-8QAM as baseline to show the nonlinearity-tolerant property of the 8D formats, and choose the 5.5b4D-2A8PSK with the same SE as baseline to show the overall performance improvement.}
 {
The  formats  were compared via two performance metrics: normalized GMI (NGMI) and effective  SNR\footnote{{The effective SNR (denoted by $\text{SNR}_{\text{eff}}$) represents the SNR after fiber propagation and the receiver digital signal processing (DSP)
and is defined as  \cite[Eq. (16)]{TobiasJLT16}.}}.   NGMI is given by NGMI=GMI/$m$, where $m$ is the number of bits per 4D of the format and shows  the  gains  for  a  BICM  system  with  the  same  soft-decision forward error correction (SD-FEC) overhead. The effective SNR quantifies the  gains  due to nonlinearity tolerance. } 

\vspace{-0.7em}
\subsection{Linear Channel Performance}\label{Sec:SimulationLinear}

\begin{figure}[!tb]
\centering
\includegraphics[scale=1]{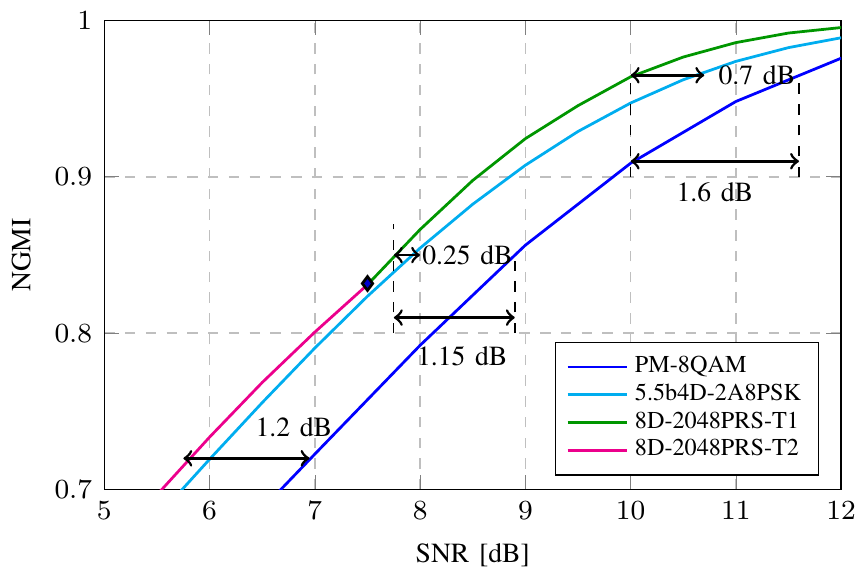}
\vspace{-2em}
  \caption{NGMI vs. SNR for linear AWGN channel.  {The black diamond
represents the switching point for two 8D formats.}}
\label{fig:NGMI_AWGN}
\vspace{-1.5em}
\end{figure}

Fig. \ref{fig:NGMI_AWGN} shows the NGMIs for the linear AWGN channel. \TypeOne and \TypeTwo are shown to clearly outperform both PM-8QAM and 5.5b4D-2A8PSK  for all NGMIs above $0.6$~bit. 
At a NGMI of 0.85 (the state-of-the-art SD-FEC with 25\%  overhead) \TypeOne offers gains of 1.15~dB and 0.25~dB with respect to PM-8QAM and 5.5b4D-2A8PSK, resp. These gains increase up to  $1.6$~dB and $0.7$~dB at high SNRs (at NGMI of 0.965~bit).

\vspace{-0.7em}
\subsection{Nonlinear Channel Performance}\label{Sec:SimulationNonLinear} 

We consider a dual-polarization multi-span WDM system with 11 co-propagating channels generated at a symbol rate of 45 GBaud,  {a WDM spacing of 50 GHz}  and a root-raised cosine
(RRC) filter roll-off factor of 0.1. Each WDM channel carries  $2^{16}$ 4D symbols in two polarizations at the same launch power per channel $P_{\text{ch}}$.   {Each span consists of an 80 km  standard single mode fiber  (SSMF) through a split-step Fourier solution of the nonlinear Manakov equation with step size 0.1~km and is  followed  by  an  erbium-doped  fiber  amplifier  with  a  noise figure  of $5$~dB. 
We also simulate polarization mode dispersion (PMD) with the coarse-step method \cite{MarcuseJLT1997}  and  fixed-length sections
of length 1~km.
As for the statistical characterisation of PMD and its effect on fiber transmission,  the polarization is uniformly scattered over Poincar$\acute{\text{e}}$ sphere and differential group delays (DGDs) of each section are   selected randomly from a Gaussian distribution with standard deviation equal to $20\%$ of the mean \cite{ProlaPTL1997}.} {At the receiver, an ideal receiver is implemented\footnote{{In this paper, we use ideal 8D phase compensation and  8D detection. However, due to the symmetric property and set-partitioned  structure of the proposed modulation family, the 8D  formats can be  equalized and demapped in 4D or even 2D with marginal loss and lower complexity \cite{BendimeradECOC2018,SjoerdECOC2019}.}} and fiber linear impairments such as the accumulated chromatic dispersion or the polarisation state rotation of the signal are \textit{ideally}\footnote{{{Ideal compensation}  refers to having at the receiver exact knowledge of the amount of  randomly generated angles and
DGD values in the fiber simulation.}} compensated.
}

\begin{figure}[!tb]
\centering
	\includegraphics[scale=1]{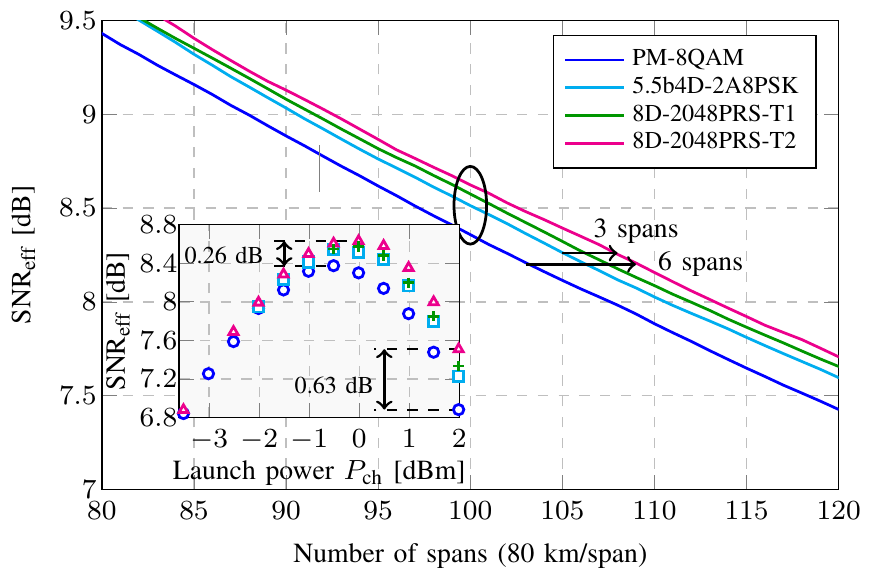}
\vspace{-2em}
	 \caption{{$\text{SNR}_{\text{eff}}$} vs. transmission distance at  $P_{\text{ch}}=0$~dBm. Inset:  {$\text{SNR}_{\text{eff}}$} vs. launch power per channel $P_{\text{ch}}$ for 8000~km link.}
	  \label{fig:effectiveSNRvsD}
\vspace{-1.5em}
\end{figure}

 {First, we consider the propagation without PMD, setting the PMD to zero.}
 {We} compare the  {$\text{SNR}_{\text{eff}}$} 
as a function of the transmission distance using $P_{\text{ch}}=0$~dBm (optimum $P_{\text{ch}}$ for 100 spans). The results are shown in Fig. \ref{fig:effectiveSNRvsD}. The two proposed 8D formats \TypeOne and \TypeTwo provide  a  higher  {$\text{SNR}_{\text{eff}}$} than PM-8QAM and 5.5b4D-2A8PSK.  
Especially, \TypeTwo has  higher SNR gains due to its smaller   nonlinearty-tolerant property of $\alpha$ and $\beta$ in Table~\ref{tab:property}. 

From the results above, we can observe that the proposed 8D-2048PRS formats outperform other modulation formats  in both linear  and nonlinear channel  {without PMD}.
The total nonlinear shaping gain is linear SNR gain (in Fig. \ref{fig:NGMI_AWGN}) plus   {$\text{SNR}_{\text{eff}}$} gain (in Fig. \ref{fig:effectiveSNRvsD}). 
 {In order to characterise the impact of the fiber PMD parameter, we consider the realistic values of the PMD in the range of $0.01-0.2~\text{ps}/\sqrt{\text{km}}$ and average the $\text{SNR}_{\text{eff}}$ over 50 random realizations of PMD for each data point.
  In Fig. 5, the average
 $\text{SNR}_{\text{eff}}$  is shown as a function of PMD  using  0~dBm  launch power per channel over a transmission distance of 8000 km. 
 The   $\text{SNR}_{\text{eff}}$  without PMD  are shown  by the dashed lines as a reference. 
Fig. 5 shows that 
the PMD has a small positive impact on the $\text{SNR}_{\text{eff}}$ for all the modulation formats.
This confirmed that random PMD depolarizes signals, averages out nonlinear effects, and thus reduces the nonlinear penalty when PMD itself is fully compensated by DSP
at receiver.
 In addition,  the average $\text{SNR}_{\text{eff}}$ gain of using 8D formats over PM-8QAM decrease  from  0.33~dB to 0.24~dB at   high PMD regime because the SOP changes during propagation.    
The inset of  Fig. 5 shows that PM-8QAM has  larger $\text{SNR}_{\text{eff}}$  variation and $0.36$~dB lower $\text{SNR}_{\text{eff}}$ in the worst-case scenario  w.r.t. 8D2048PRS-T2.  
}

\begin{figure}[!tb]
\centering
\includegraphics[scale=1]{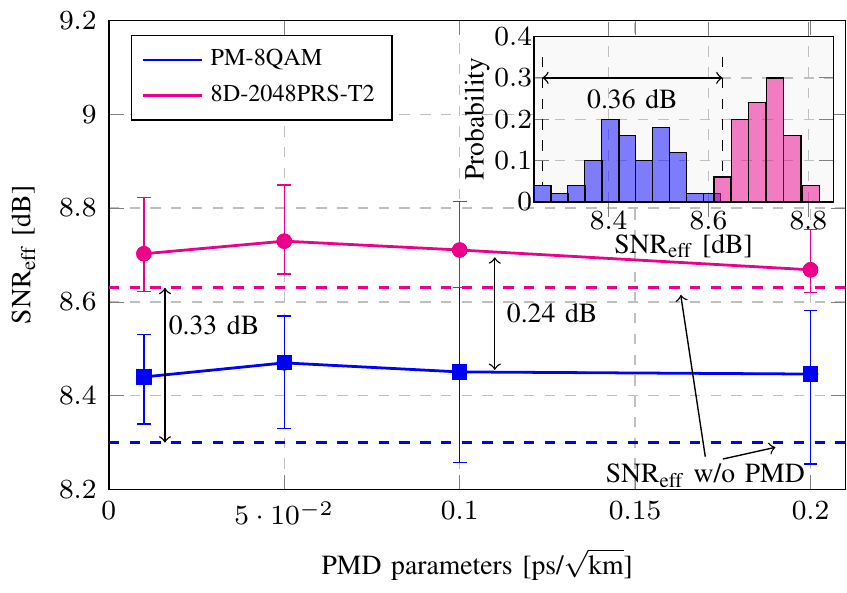}
	 \vspace{-1em}
	\caption{{Average $\text{SNR}_{\text{eff}}$  as a function of the fiber PMD parameter at $P_{\text{ch}}=0$~dBm for  transmission over 8000 km. Inset: Histograms of $\text{SNR}_{\text{eff}}$  values obtained  for PM-8QAM and 8D-2048PRS-T2 with  PMD=0.1~ps/$\sqrt{\text{km}}$.}}
	  \label{fig:NGMIvsPMD}
\vspace{-1em}
\end{figure}

\begin{figure}[!tb]
\centering
\includegraphics[scale=1]{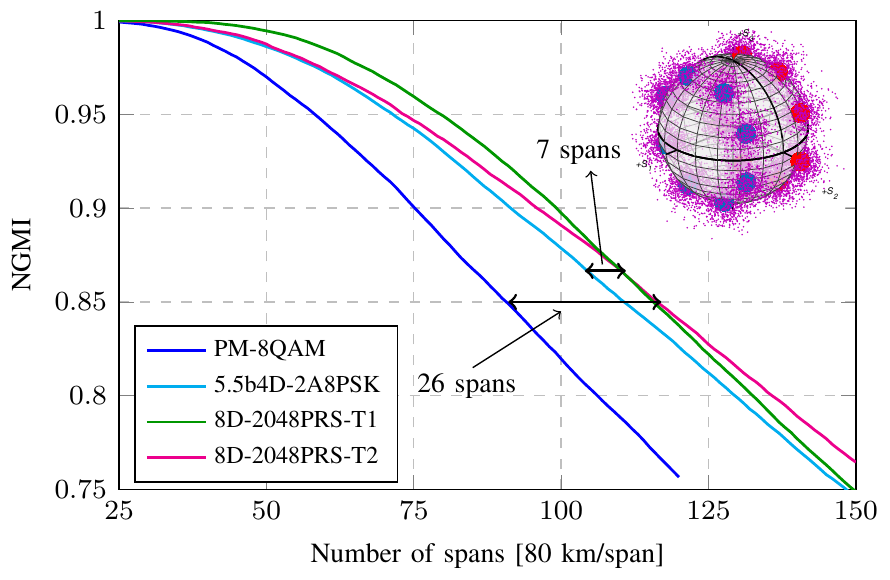}
	 \vspace{-2em}
	 \caption{NGMI versus transmission distance  {(without PMD)}. Inset: Stoke space projection of the received symbols for \TypeTwo after 20 spans. 
	 }
	  \label{fig:GMIvsD}
\vspace{-1em}
\end{figure}

Fig. \ref{fig:GMIvsD} shows the results  {without PMD} of the NGMI  {as  a  function} of the transmission distance, using the optimal launch power at each distance. In addition, the recovered \TypeTwo constellation after 20 spans (in Stokes space) is inset.  {Note that both proposed constellations yield  a 26 spans ($28.6\%$) and 7 spans ($6.7\%$)  reach increase   relative  to PM-8QAM and 5.5b4D-2A8PSK at NGMI of 0.85. }

\vspace{-0.5em}
\section{Conclusions}
We have designed two new nonlinearity-tolerant 8D modulation formats at spectral efficiency of 5.5 bits/4D-sym 
 {and have provided a simple bit-to-symbol mapping by set-partitioning, which shows that these format can be implemented with slight modifications to 4D-64PRS.  {Although at a lower SE, the 8D-2048PRS formats outperforms PM-8QAM by significantly improving sensitivity and nonlinearity tolerance.}
In comparison to modulation formats of the same spectral efficiency such as 5.5b4D-2A8PSK, $6.7\%$ reach increase is observed. 
The impact of PMD on the proposed modulations were numerically analyzed to show tolerance to nonlinear effects. We believe that the proposed 8D formats are promising candidates for transmission systems with high nonlinearity, and can be extended to higher dimensions, including wavelengths, and mode/core spatial channels.  {Future work will also address a realistic comparison with probabilistic shaping.}}

\small{\noindent \textbf{Acknowledgements:} The authors would like to thank Dr. Gabriele Liga (Eindhoven University of Technology, The Netherlands) for the useful discussions.}
\vspace{-1em}
\bibliographystyle{IEEEtran}
\bibliography{references}

\end{document}

%% file: myFigures.tex
\usetikzlibrary{plotmarks,matrix,chains,scopes,fit,calc,shapes,positioning,decorations,intersections,arrows,backgrounds,shadows}

\newlength\FigureHeight
\newlength\FigureWidth


%% file: myStyles.tex
\definecolor{myDarkGreen}{rgb}{0.00000,0.58824,0.00000}%
\definecolor{uniform}{rgb}{0.00000,0.58824,0.00000}%
\definecolor{AWGNreference}{rgb}{0.00000,0.58824,0.00000}%

%% file: myMacros.tex
\newcounter{lemma}

\newtheorem{exampleplain}{Example}

\newcommand{\bS}{\boldsymbol{S}}



%% file: myPackages.tex
\usepackage[utf8]{inputenc}

\usepackage{amsmath,amssymb}

\usepackage{xcolor}
\usepackage{graphicx}
\usepackage{tikz}
\usepackage{pgfplots}
\usepackage{float}
\usepackage{balance}

\usepackage{xspace}
\usepackage[normalem]{ulem}
\usepackage[pdftex]{hyperref} 

\usepackage{dsfont,empheq}
\usepackage{multicol,multirow}

\usepackage{cite}
\usepackage{gensymb}